\documentclass[aps,prb,twocolumn,groupedaddress,showpacs]{revtex4-1}
\usepackage{amsmath}
\usepackage{graphicx}
\usepackage{epstopdf}

\newcommand{\TC}{ \mathrm{TC} }
\newcommand{\RK}{ \mathrm{RK} }

\newcommand{\ket}[1]{ \vert #1 \rangle}

\newcommand{\eket}[1]{ \left \vert #1 \right \rangle}
\newcommand{\ebra}[1]{ \left \langle #1 \right \vert}

\newcommand{\eref}[1]{Eq. (\ref{#1})}
\newcommand{\figref}[1]{Fig. \ref{#1}}
\newcommand{\plaqa}{
 {\mathchoice
  {\includegraphics[height=1.6ex]{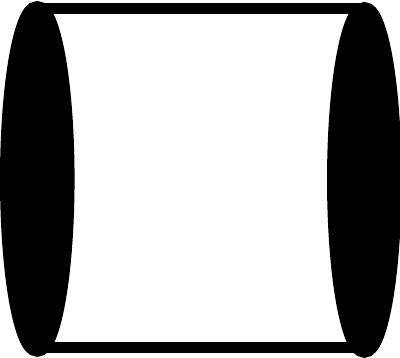}}
  {\includegraphics[height=1.6ex]{plaqa}}
  {\includegraphics[height=1.2ex]{plaqa}}
  {\includegraphics[height=0.9ex]{plaqa}}
 }
}
\newcommand{\plaqb}{
 {\mathchoice
  {\includegraphics[height=1.6ex]{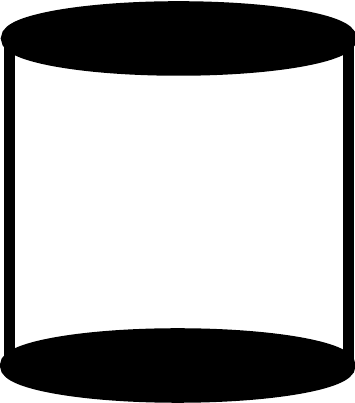}}
  {\includegraphics[height=1.6ex]{plaqb}}
  {\includegraphics[height=1.2ex]{plaqb}}
  {\includegraphics[height=0.9ex]{plaqb}}
 }
}

\begin{document}

\title{Geometric properties of loop condensed phases on the square lattice}
\author{C. M. Herdman$^{1,2}$}
\altaffiliation[Present Address: ]{Department of Physics, University of Vermont, Burlington, Vermont 05405, USA}
\email{Christopher.Herdman@uvm.edu}
\author{K. B. Whaley$^2$}
\affiliation{Berkeley Center for Quantum Information and Computation, Departments of Physics$^{1}$ and Chemistry$^{2}$, University of California, Berkeley, California 94720, USA}
\date{\today}

%----------------------------------------------------------------
%------   Abstract
%----------------------------------------------------------------
\begin{abstract}
Loop condensed phases are scale-invariant quantum liquid phases of matter. These phases include topologically ordered liquid phases such as the toric code as well as critical liquids such as the Rokhsar-Kivelson point of the quantum dimer model on the square lattice. To investigate the extent to which nonlocal geometric observables capture a signature of the nonlocal quantum order present in these phases, we compute geometric properties of such loop condensed states using directed loop Monte Carlo calculations. In particular, we investigate the loop condensed nature of ground state of the square lattice quantum dimer model at the Rokhsar-Kivelson point and compare with other loop condensed states on the square lattice, including those of the toric code and fully packed loop model. The common features of such liquids are a scale invariant distribution of loops and a fractal dimensionality of spanning loops. We find that the fractal dimension of the loop condensate of the square lattice quantum dimer model at the Rokhsar-Kivelson point is $3/2$, which provides quantitative confirmation of the effective height model that is commonly used to describe this critical dimer liquid.
\end{abstract}


\maketitle

%----------------------------------------------------------------
%------Introduction  
%----------------------------------------------------------------
\section{Introduction}

The discovery of quantum liquid phases of matter that possess quantum order without breaking conventional symmetries has demonstrated that ordered quantum phases exist outside of Landau's paradigm of conventional phases of matter. To understand the formation of order in such quantum liquid phases, the mechanism of loop and string-net condensation has been proposed by several authors ~\cite{Kitaev2003,Levin2005a,Fendley2005,Freedman2004,Fendley2008}. In this description of quantum liquids, the effective degrees of freedom are extended objects (loops or string-nets) that fluctuate on all length scales and generate the nonlocal quantum entanglement present in such phases. In particular, loop condensates can describe $Z_2$ topologically ordered phases as well as critical quantum liquids~\cite{Levin2005a,Freedman2004}.

Since these quantum liquids cannot be described by a local order parameter, the presence of order in these phases must be characterized by other quantities. For example, signatures of topological order appear in the bipartite entanglement entropy of a topologically ordered quantum liquid~\cite{Levin2006a,Kitaev2006b}. A relevant question is whether these quantum orders have signatures in more conventional observables.  To make direct contact with the loop condensate picture of a quantum liquid, we seek to understand to what extent quantum liquid phases can be characterized directly in terms of the effective loop degrees of freedom. For example, in addition to local loop correlation functions, loop condensates can be described by the fractal dimension of the loop gas~\cite{Troyer2008}.

Lattice models that generate local constraints on the ground state often possess loop condensed liquid phases. The toric code~\cite{Kitaev2003} ground state can be described as having a local constraint requiring that an even number of loop segments touch each vertex~\cite{Kitaev2003,Trebst2007}. The $Z_2$ topologically ordered ground state is simply the loop condensed phase of this (intersecting) closed loop subspace. Other models with local constraints may not explicitly be loop models, but the local constraint often may be readily mapped to a closed loop constraint. For example, quantum dimer models~\cite{Moessner2008,Rokhsar1988} impose a hard constraint at each vertex requiring a single dimer to touch each vertex; however by superimposing dimer configurations on a reference dimer configuration, a dimerization can be mapped onto a loop covering of the lattice~\cite{Sutherland1988b,Sutherland1988,Sutherland1988a,Kohmoto1988a,Kohmoto1988}. On the square lattice, the quantum dimer model has an isolated critical dimer liquid point~\cite{Rokhsar1988,Leung1996,Syljuasen2006}, but on the triangular lattice there is a gapped topologically ordered liquid phase~\cite{Moessner2001,Ioselevich2002a,Fendley2002,Ralko2005a}. In this work and in Ref. \citenum{Herdman2011}, we show that these dimer liquid phases can be viewed as loop condensed phases~\cite{Herdman2011}. 

In this paper, we investigate loop condensed states that arise in locally constrained models on the square lattice. We explicitly demonstrate the loop condensed nature of several liquid states in geometrically constrained models on the square lattice, including the square lattice quantum dimer model. In particular, we compute the fractal dimension of the underlying loop gas, and examine the fractal dimension as a distinguishing feature of these phases. We find that the fractal dimension of the critical dimer liquid at the Rokhsar-Kivelson (RK) point of the square lattice dimer model is $3/2$. This agrees with that of the contour loops of the effective height model description of the RK point~\cite{Kondev1995}, providing quantitative validation of the height model.

%----------------------------------------------------------------
%------Introduction to Loop condensates
%----------------------------------------------------------------
\section{Introduction to Loop condensates}

A loop condensate~\cite{Fendley2008,Freedman2004,Fendley2005,Levin2005a} is a scale invariant liquid state with fluctuating loops on all length scales. We may describe such a loop liquid with both conventional local correlation functions and non-local geometric observables. Local correlation functions include the loop-loop correlation function and the defect-defect correlation function. The loop-loop correlation function determines the probability that loop segments exist at two points separated by $r$. This distinguishes between purely disordered loop liquids with exponentially decaying local correlations, and those with quasi-long-range order due to local correlations that decay as a power law. The defect-defect correlation function determines the correlations between defects that live at the end of a broken loop separated by a distance $r$. 
While these local correlation functions display the disordered nature of loop liquids, they do not capture the nature of nonlocal quantum order present in these phases.

Nonlocal, geometric properties of loop condensates are described by the loop distribution $P(s)$, and the fractal dimension $D_f$, which are related to the two-loop correlation function $G_2(r)$. Scale invariance demands that the loop distribution, $P(s)$, the probability that a loop has length $s$, scales as a power law of length:
\begin{align}
P\left(s\right) \sim s^{-\tau}. \label{Ps}
\end{align}
Additionally, the scaling of the length of a loop $\ell$ with its radius $R(\ell)$ determines the fractal dimension $D_f$ of the loop condensate:
\begin{align}
s\left(\ell\right) \sim R\left(\ell\right)^{D_f}.
\end{align}
These geometric exponents may be related by a scaling relation to the two-loop correlation function $G_2(r)$ that determines the probability that two points separated by a distance $r$ lie on the same loop~\cite{Saleur1987}. Since the two-loop correlation function is fundamentally nonlocal (since it is determined by the existence of a single loop connecting the points), $G_2(r)$ may decay as a power law even in the absence of local correlations. The exponent $x_2$ that governs the power-law decay $G_2\left(r \right) \sim {r^{-2x_2}}$ determines both the geometric exponents $\tau$ and $D_f$~\cite{Saleur1987,Kondev1995}:
\begin{align}
D_f = 2-x_2,\quad \tau = 1+\frac{2}{2-x_2}. \label{scaling}
\end{align}
In the remainder of this paper we will investigate to what extent nonlocal, geometric loop observables such as the loop distribution and the fractal dimension display signatures of the quantum order of loop condensates on the square lattice.

%----------------------------------------------------------------
%------Loop condensation from local constraints
%----------------------------------------------------------------
\section{Loop condensation from local constraints}

\begin{figure}[] 
\centering
\includegraphics[width=2.5in]{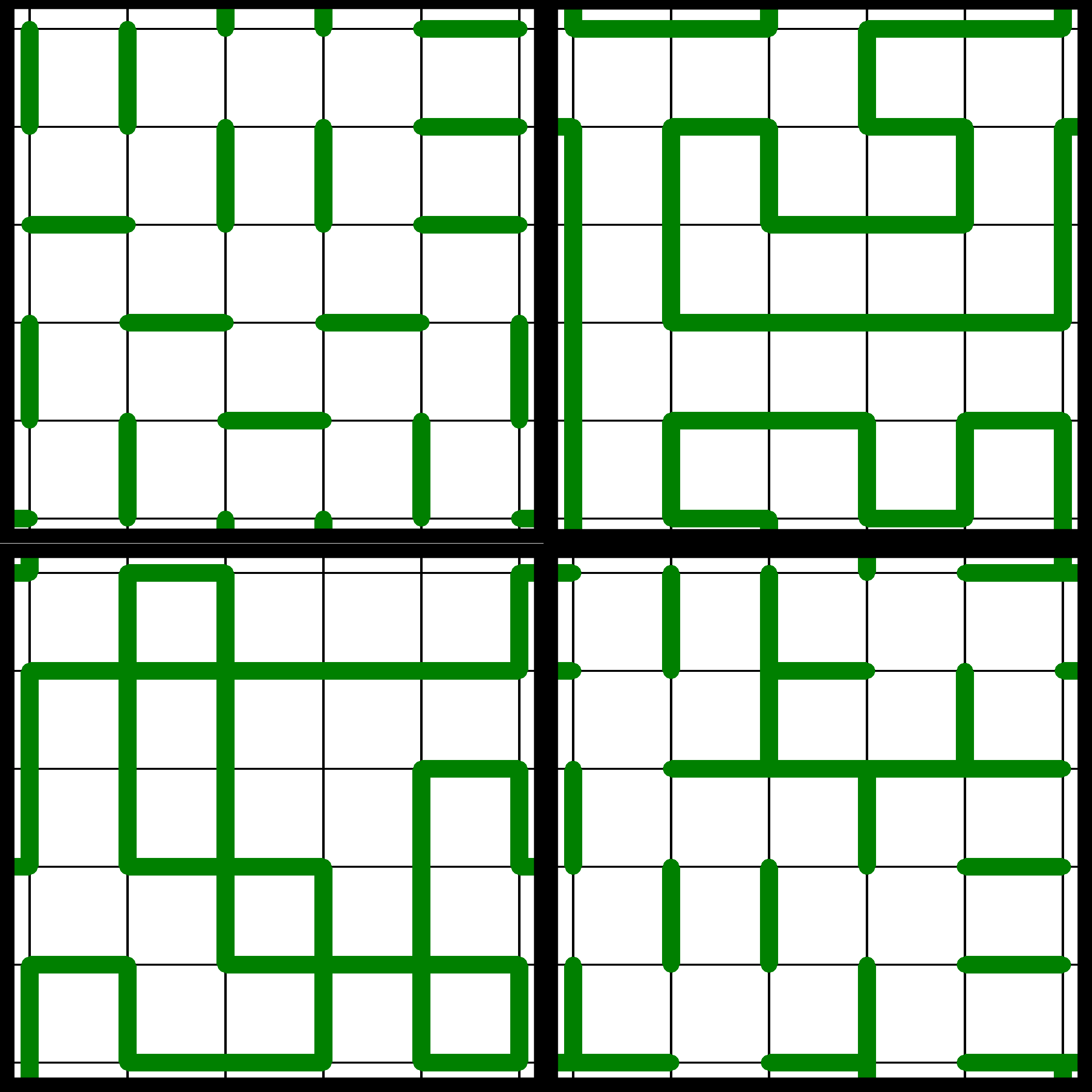} 
 \caption{ 
Examples of configurations with the local constraints on the square lattice. Clockwise from upper left: $\{C_1\}$, $\{C_2\}$, $\{C_o\}$, $\{C_e\}$, which have local constraints that restrict configurations to have a one, two, an odd and an even number of dimers touching each vertex, respectively.
}
\label{fig_sq_constraint}
\end{figure}

Lattice models with local constraints can often be mapped to loop models, and consequently display a variety of loop condensed phases. Consider a lattice model with Ising degrees of freedom that live on the links of the lattice: the two states of each Ising degree of freedom may be considered to correspond to a link that is either occupied or unoccupied by a dimer. A link $l$ (un)occupied by a dimer is the $+1$($0$) eigenstate of the dimer number operator $n_l = d_l^+ d_l^-$, with $d_l^{+/-}$ the dimer creation and annihilation operators. The simplest local constraint is to fix the dimer number at each vertex to be a constant $n_0$; we define the configuration space $\{C_{n_0}\}$ to be the set of configurations with $n_0$ dimers touching each vertex (see Fig. \ref{fig_sq_constraint}). On a square lattice there are only two distinct non-trivial cases: $n_0 = 1$ (or equivalently $n_0 = 3$) and $n_0 = 2$. $\{C_1\}$ comprises fully packed, hard-core dimerizations of the lattice. $\{C_2\}$ comprises fully packed, nonintersecting loop configurations with two dimers touching each vertex.

All fully packed hard-core dimerizations in $\{C_1\}$ can be mapped to a non-intersecting closed loop configuration by choosing a reference dimerization $R_0$ (see Fig. \ref{fig_sq_height}). When a dimerization is superimposed over $R_0$, every vertex will be touched by one physical dimer and by one reference dimer. With the exception of links where a physical dimer coincides with a reference dimer, the combined physical and reference dimers form a closed loop covering of the lattice, where the loops comprise sequences of links that are alternately occupied by physical and reference dimers, respectively. We may choose to define links with overlapping physical and reference dimers to be unoccupied in the loop picture--with this choice the physical dimer configuration that completely coincides with $R_0$ corresponds to the empty loop configuration.

The simplest liquid state in these constrained Hilbert spaces is given by a generalization of the Rokhsar-Kivelson wave function of the quantum dimer model~\cite{Rokhsar1988}. This generalized Rokhsar-Kivelson wave function $\ket{\Psi_\RK^{n_0}}$ is an equal super-position of all configurations obeying the constraint:
\begin{align}
\eket{\Psi_\RK^{n_0}} \equiv \frac{1}{\sqrt{\mathcal{N}_{n_0}}} \sum_{C_{n_0}} \eket{C_{n_0}}. \label{RK}
\end{align}
On a surface with trivial topology, the sum in (\ref{RK}) is over all configurations $C_{n_0}$ that obey the given local constraint and $\mathcal{N}_{n_0}$ is the number of such configurations. For $n_0 = 1$, $\ket{\Psi_\RK^1}$ is the Rokhsar-Kivelson state of the quantum dimer model. Diagonal expectation values of $\ket{\Psi_\RK^{n_0}}$ are equal to those of the corresponding classical model~\cite{Rokhsar1988}. Consequently, the behavior of correlation functions and local order parameters of $\ket{\Psi_\RK^{n_0}}$ are determined by the statistical mechanics of the related classical models. Diagonal correlation functions of $\ket{\Psi_\RK^{1}}$ are determined by the classical dimer model, and those of $\ket{\Psi_\RK^{2}}$ by the classical fully-packed loop model, with a loop fugacity equal to $1$ (which maps to the $6$-vertex model)~\cite{Jacobsen1998,Baxter1970}.

Rokhsar and Kivelson defined a local Hamiltonian for which $\ket{\Psi_\RK^{n_0}}$ is the exact zero-energy ground state at one point in the parameter space~\cite{Rokhsar1988}:
\begin{align}
H_\RK = -t \sum_p \Bigl( \eket{\plaqa}\ebra{\plaqb} + \mathrm{h.c.} \Bigr) + v \sum_p \Bigl( \eket{\plaqa}\ebra{\plaqa} +\eket{\plaqb}\ebra{\plaqb} \Bigr). \label{Hrk}
\end{align}
In \eref{Hrk}, the sum is over all plaquettes $p$ of the lattice. The ``$t$" term flips dimers around a ``flippable" plaquette (i.e., a plaquette with alternating occupied and unoccupied links) and the ``$v$" term is a potential energy for such flippable plaquettes. For $v=t$, the so-called RK point, $\ket{\Psi_\RK^{n_0}}$ is the zero energy ground state of $H_\RK$. We note that $H_\RK$ is specifically  constructed such that it conserves the dimer number at a given vertex, and therefore does not violate the local dimer number constraint. 

On a surface of nontrivial topology, such as a torus, the local dynamics of $H_\RK$ will break $\{C_{n_0}\}$ into distinct topological sectors. $H_\RK$ will not connect configurations in different topological sectors and thus generates a topological degeneracy at the RK point. Indeed, a distinct ground state $\ket{\Psi_\RK^{n_0}}$ can be defined within each topological sector by limiting the sum in \eref{RK} to configurations within the topological sector. The nature of this topological degeneracy is manifest in the loop condensate picture: each topological sector corresponds to a distinct winding sector defined by the number of loops winding around each topologically nontrivial cycle of the surface~\cite{Rokhsar1988}.

\textcite{Moessner2001} have shown that a local dimer number constraint generates a local $U(1)$ invariance. If we define the operator $n_v = \sum_{l \in v} n_l$ where $n_l$ is the dimer number operator on the link $l$, then all states $\ket{\psi_{n_0}}$ that obey the local constraint are invariant under the following local gauge transformation:
\begin{align}
G^{n_0}_v \left( \alpha \right) \equiv \exp{ \Bigl( i \alpha \left( n_v - n_0 \right) \Bigr)}, \quad G^{n_0}_v \left( \alpha \right) \eket{\psi_{n_0}} = \eket{\psi_{n_0}}.
\end{align}
%This suggests that if $\ket{\Psi_\RK^{n_0}}$ is in a liquid phase, $H_\RK$ may be related to a $U(1)$ gauge theory.
Consequently, $H_\RK$ has been shown to map to a $U(1)$ gauge theory for both $n_0 = 1$~\cite{Fradkin1990,Fradkin1998} and $n_0 = 2$~\cite{Ardonne2004}.

We now consider relaxing the dimer number constraint to a local dimer parity constraint. Here, $\{C_e\}$ and $\{C_o\}$, which comprise configurations with a fixed dimer number parity at each vertex, have an even or odd number of dimers touching each vertex, respectively. The physical states $\ket{\psi_{e,o}}$ that are formed from superpositions of $\{\ket{C_{e,o}}\}$ are invariant under the gauge transformations
\begin{align}
G^{e}_v \equiv \exp{ \bigl( \pm i \pi n_v \bigr) }, \quad &G^{e}_v  \eket{\psi_e} = \eket{\psi_e} \\
G^{o}_v \equiv \exp{ \Bigl( \pm i \pi \left( n_v +1\right) \Bigr) }, \quad &G^{o}_v \eket{\psi_o} = \eket{\psi_o}.
\end{align}
Note that the $U(1)$ symmetry of $G^{n_0}_v \left( \alpha \right)$ has thereby been reduced to a $Z_2$ symmetry in $G^{e,o}_v$\cite{Moessner2001}.

We may now define a Hamiltonian that commutes with $G^{e,o}_v$:
\begin{align}
H_{\mathrm{TC}} = -t \sum_p \prod_{l \in p} d_l^x, \label{Htc}
\end{align}
where $d_l^x \equiv 1/\sqrt{2} (d_l^+ + d_l^-)$ and $d_l^{+/-}$. This is the magnetic term of the toric code Hamiltonian\cite{Kitaev2003}. In the toric code Hamiltonian, the dimer parity constraint is imposed by a local energy cost at each vertex; here, we take that energy cost to be infinite, such that this is a hard constraint. The ground state of \eref{Htc} is the equal superposition of all configurations in the given parity sector:
\begin{align}
\eket{\Psi_\TC^{e,o}} \equiv \frac{1}{\sqrt{\mathcal{N}_{e,o}}} \sum_{C_{e,o}} \eket{C_{e,o}}. \label{TC}
\end{align}
In the language of the \textcite{Moessner2001}, $\ket{\Psi_\TC^{e,o}}$ in the even (odd) parity sector corresponds to the even (odd) Ising gauge theory; $\ket{\Psi_\TC^{e}}$ is simply the toric code ground state.

We may now understand the relationship between the $U(1)$ gauge theory describing \eref{RK} with a local dimer number constraint, and the $Z_2$ gauge theory that describes \eref{TC} with a fixed dimer parity constraint. Since the square lattice is bipartite, the links can be oriented to always point from one sublattice to the other; dimers can be viewed as carrying a flux into or out of a vertex according to this orientation. The $n_v = 2$ constraint acts as Gauss's law, such that there is a flux of $+2$ and $-2$ on each vertex of sublattices A and B respectively; correspondingly we can consider there to be a static background charge of $\pm2$ on the two sublattices. Now consider the case when $\ket{\Psi_\RK^{2}}$ is doped with dynamic $n_v =0,4$ vertices; these will act as charge $\pm2$ objects, depending on the sublattice. Allowing $n_v =0,4$ vertices will transform $\ket{\Psi_\RK^{2}}$ into $\ket{\Psi_\TC^{e}}$, which is the toric code ground state, and is known to be described by a $Z_2$ gauge theory~\cite{Kitaev2003}. 

This transition from a $U(1)$ gauge theory to a $Z_2$ gauge theory via the introduction of charge $2$ objects follows the well-known prescription of Fradkin and Shenker, who showed that coupling a $U(1)$ gauge field to charge $N>1$ matter field can reduce the $U(1)$ gauge symmetry down to $Z_N$~\cite{Fradkin1979}. The same picture applies in the odd parity sector, where the QDM can be viewed as a $U(1)$ gauge theory with $\pm1$ background charges on the sublattices. In this case, introducing $n_v=3$ vertices is equivalent to allowing charge 2 matter fields and it generates $\ket{\Psi_\TC^{o}}$, which is described by the odd Ising gauge theory. This construction leads one naturally to ask whether other $Z_N$ models may live at the RK points of geometrically constrained models on the square lattice. In particular, doping the QDM with charge $\pm3$, $n_v=4$ vertices might lead to a $Z_3$ liquid phase, purely as a result of geometrical constraints.

%%%%%%%%%%%%%%%%%%%%%%%%%
%Square lattice quantum dimer model
%%%%%%%%%%%%%%%%%%%%%%%%%
\section{Square lattice quantum dimer model} 

\begin{figure}[] 
\centering
\includegraphics[width=2in]{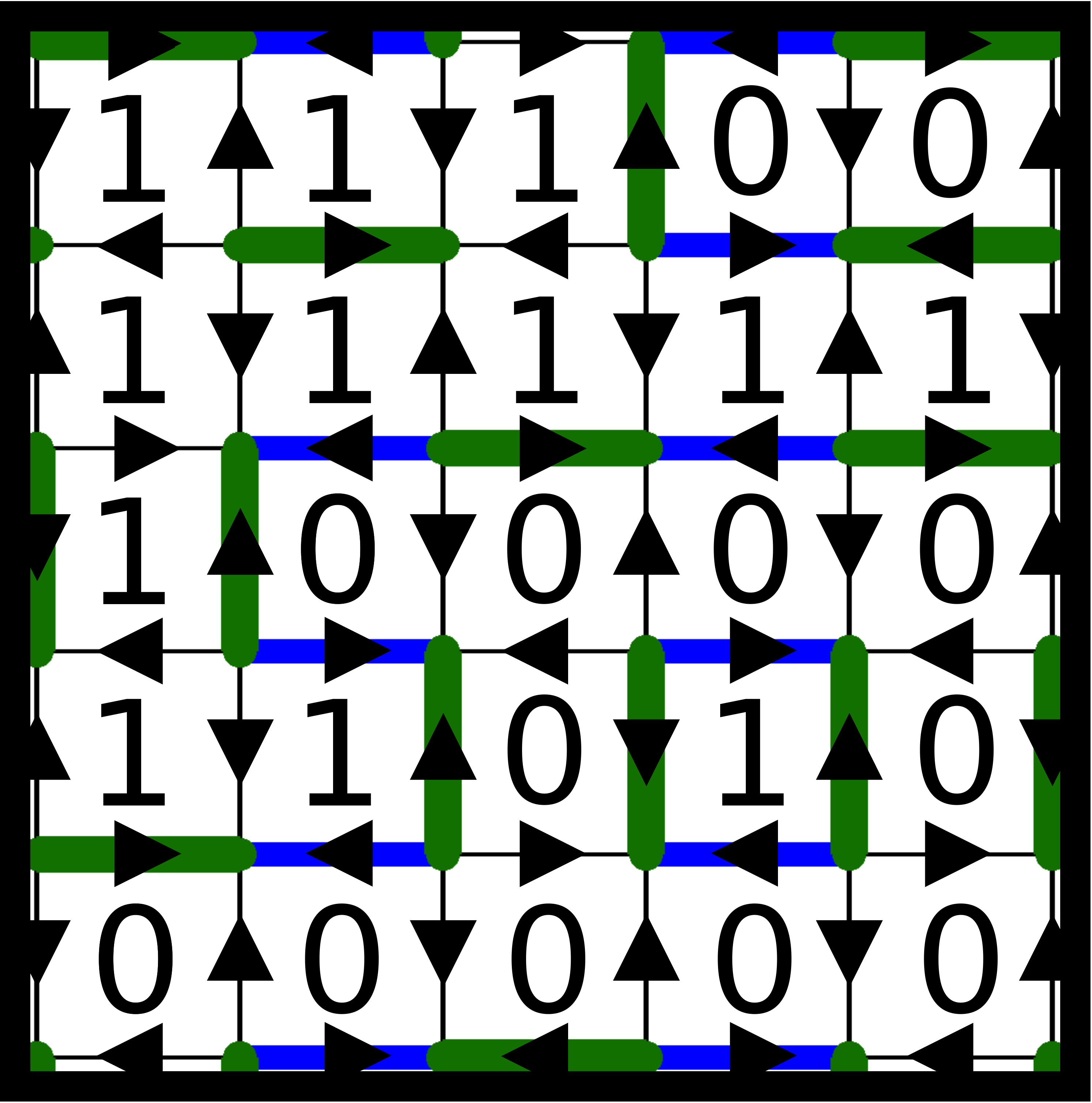} 
 \caption{ 
An example of the mapping of the square lattice QDM to a height  model. The green dimers represent the physical dimers and the blue dimers are the background dimerization. The lower left plaquette is chosen to have $h_0=0$ and all other heights are determined relative to this, by following a path from $h_0$ through the links of the lattice. If a dimer is crossed with the arrow pointing to the right (left), the height changes by $+1$ ($-1$) and vice versa for a reference dimer. Flipping dimers on a flippable plaquette changes the local height of the plaqeutte by $\pm1$. The transition loops formed by alternating green and blue dimers are the contour loops of the height field.
}
\label{fig_sq_height}
\end{figure}

The ground state of the RK point of the square lattice QDM is a gapless critical dimer liquid, with power-law decaying dimer-dimer correlations~\cite{Rokhsar1988}.
As described above, a dimer model may be mapped to a closed loop model by introducing a reference dimerization. As such, we may view the RK point of the square lattice QDM as a loop condensed liquid phase. There is a well known phenomenological height model description of the square lattice RK wave function~\cite{Henley1997,Kondev1996a,Kondev1997,Raghavan1997}. Here we will give a variant of this height model picture that makes the meaning of the loop picture more transparent (see Fig. \ref{fig_sq_height}). (1) First, we orient the links such that the arrows point from sublattice A to B and choose a reference plaquette to assign a height of $h=0$, as in figure \ref{fig_sq_height}. (2) Heights $h(r)$ are assigned to all other plaquettes by starting with the reference plaquette and moving along any path crossing the links, where we assign a $\delta h$ for each link, with $\vert \delta h \vert = 1$. (3) If a link crossed with the arrow pointing to the right is occupied by a \emph{physical} dimer,  then $\delta h = +1$, if it is occupied by a \emph{reference} dimer then $\delta h = -1$, and vice versa if the link points to the left. (4) Flipping parallel dimers on a plaquette will change the local height by $\pm1$. With this mapping, the transition loops formed from the combination of the reference and physical dimerizations act as the contour loops of the height field, since the height only changes when a transition loop is crossed (see \figref{fig_sq_height}).

We then propose an effective Gaussian action $S_\RK$ for a coarse grained height field that captures the local height fluctuations and describes the dimer liquid at the RK point:
\begin{align}
S_{\RK} = \int dr^2 \frac{K}{2} \left \vert \bigtriangledown h \right \vert^2 + V\left(h\right).\label{Sheight}
\end{align}
The statistical weight for a given configuration of heights is then proportional to $exp(-S_{\RK})$. In \eref{Sheight}, $K$ is the stiffness and the first term captures the fluctuations of the height field. $V(h)$ is a locking potential that favors certain ordered height configurations. At the RK point on the square lattice, the locking potential is irrelevant in the renormalization group sense, and therefore the effective action is Gaussian. This Gaussian action describes a rough phase of the height model, where $h$ fluctuates along the lattice. Such a phase can be shown to have power law correlation functions~\cite{Henley1997,Kondev1996a,Kondev1997,Raghavan1997}. Consequently, the height model can capture the critical correlations of the dimer liquid at the RK point, with an appropriate choice of the stiffness $K$. We note that this height model description is purely phenomenological; currently, there is no well known microscopic derivation of this action for the QDM. 

The transition loops to the reference dimerization take on a special meaning in the height model. In particular they are the contour loops of the height model, in that the height only changes when a transition loop is crossed (see \figref{fig_sq_height}). The fractal dimension of the contour loops of a Gaussian height model has been shown to be universal (independent of the stiffness $K$) and equal to $3/2$ by Kondev and Henley~\cite{Kondev1995}. 

\begin{figure}[h] 
\centering
 \includegraphics[width=3.5in]{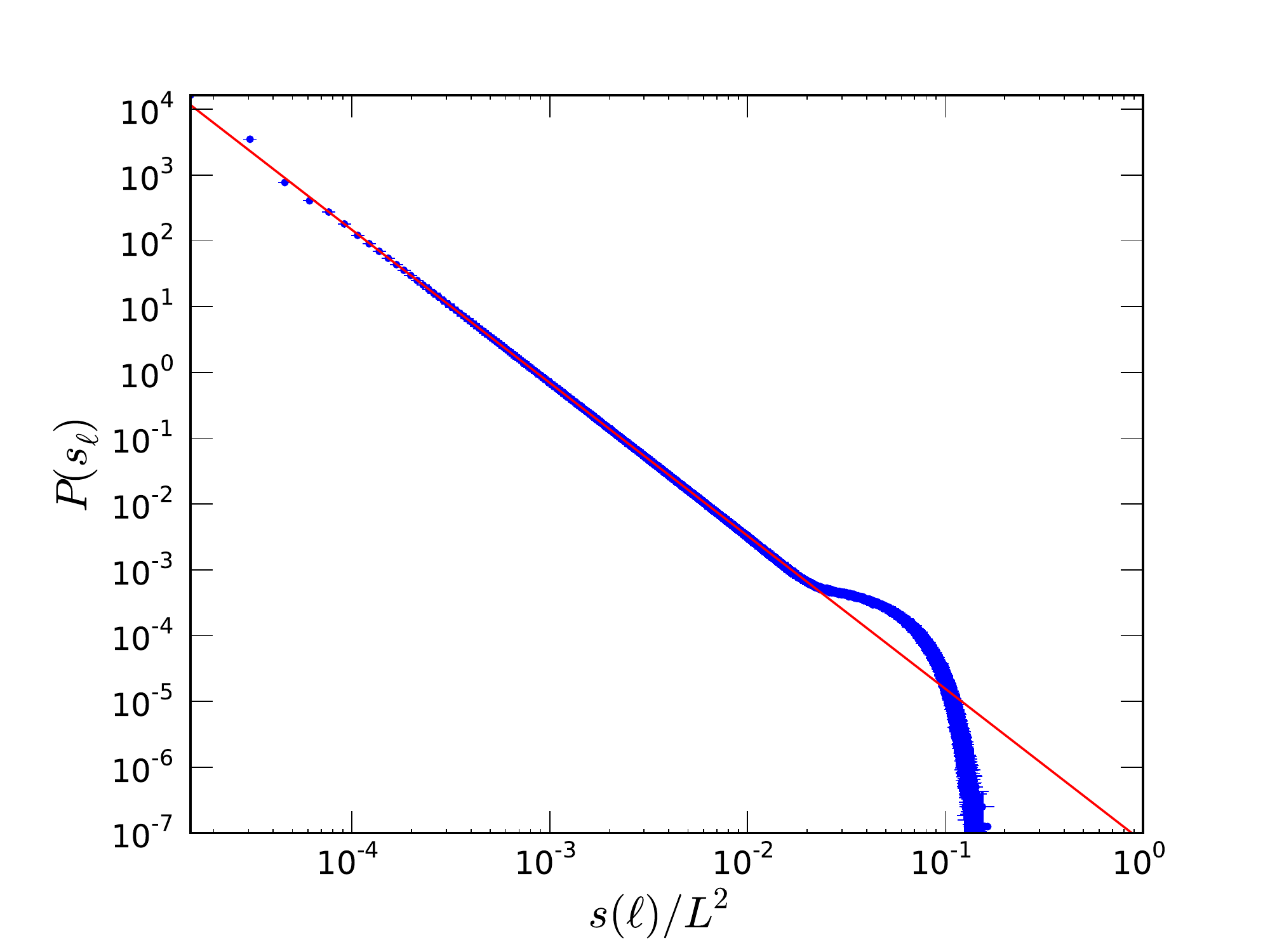} 
 \caption{ 
The loop distribution function $P\left(s\right)$ for the square lattice dimer model at the RK point for a lattice of linear dimension $L=512$. $P(s)$ displays a clear power law over length scales $s \ll L^2$. The line shows the best fit power law with $\tau=2.32\pm0.02$.
}
\label{fig_sqRK_svl}
\end{figure}

We have computed the distribution of the transition loops of the square lattice QDM at the RK point with directed loop Monte Carlo calculations~\cite{Syljuasen2004,Sandvik2006}. We first consider the computed loop distribution shown in \figref{fig_sqRK_svl}. We see that $P(s)$ displays a power law behavior over loops on length scales $s<<L^2$, which is indicative of the loop condensed nature of the dimer liquid at the RK point. The best fit power law gives $\tau = 2.32\pm0.02$. In the effective height model, $x_2 = 1/2$ and [from the scaling relation given in \eref{scaling}] $\tau=7/3$, so our result is quantitatively consistent with the predictions of height model description.

\begin{figure}[h] 
\centering
 \includegraphics[width=3.5in]{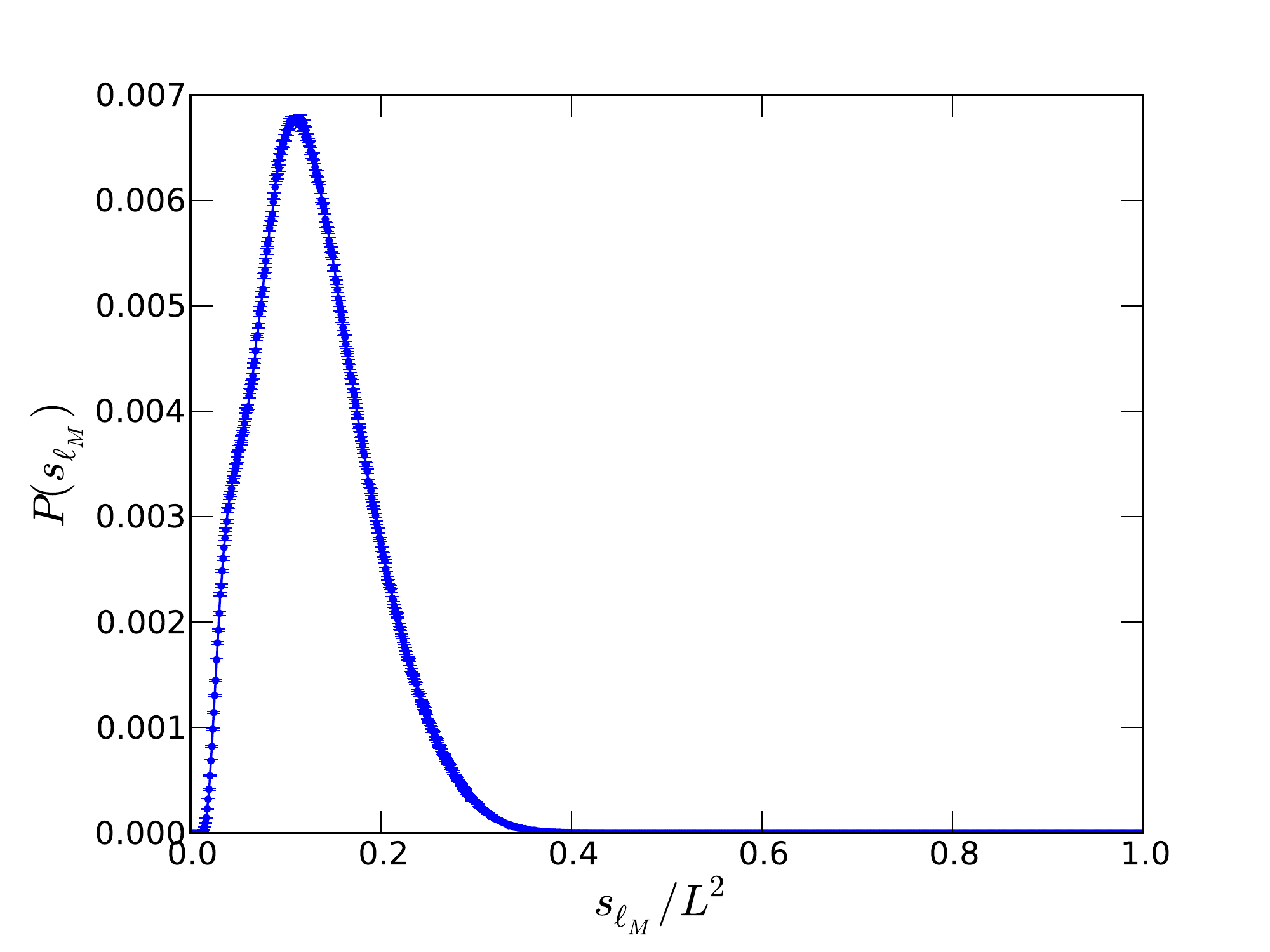} 
 \caption{ 
The distribution of the longest loops , $P\left(s_M \right)$ for the square lattice quantum dimer model at the RK point on an $L=64$ lattice. The broad distribution over many length scales is indicative of the loop condensed nature of the state. The location of the peak scales with system size as shown in Fig. \ref{fig_sqRK_svL}.
}
\label{fig_sqRK_ll}
\end{figure}

\begin{figure}[h] 
\centering
 \includegraphics[width=3.5in]{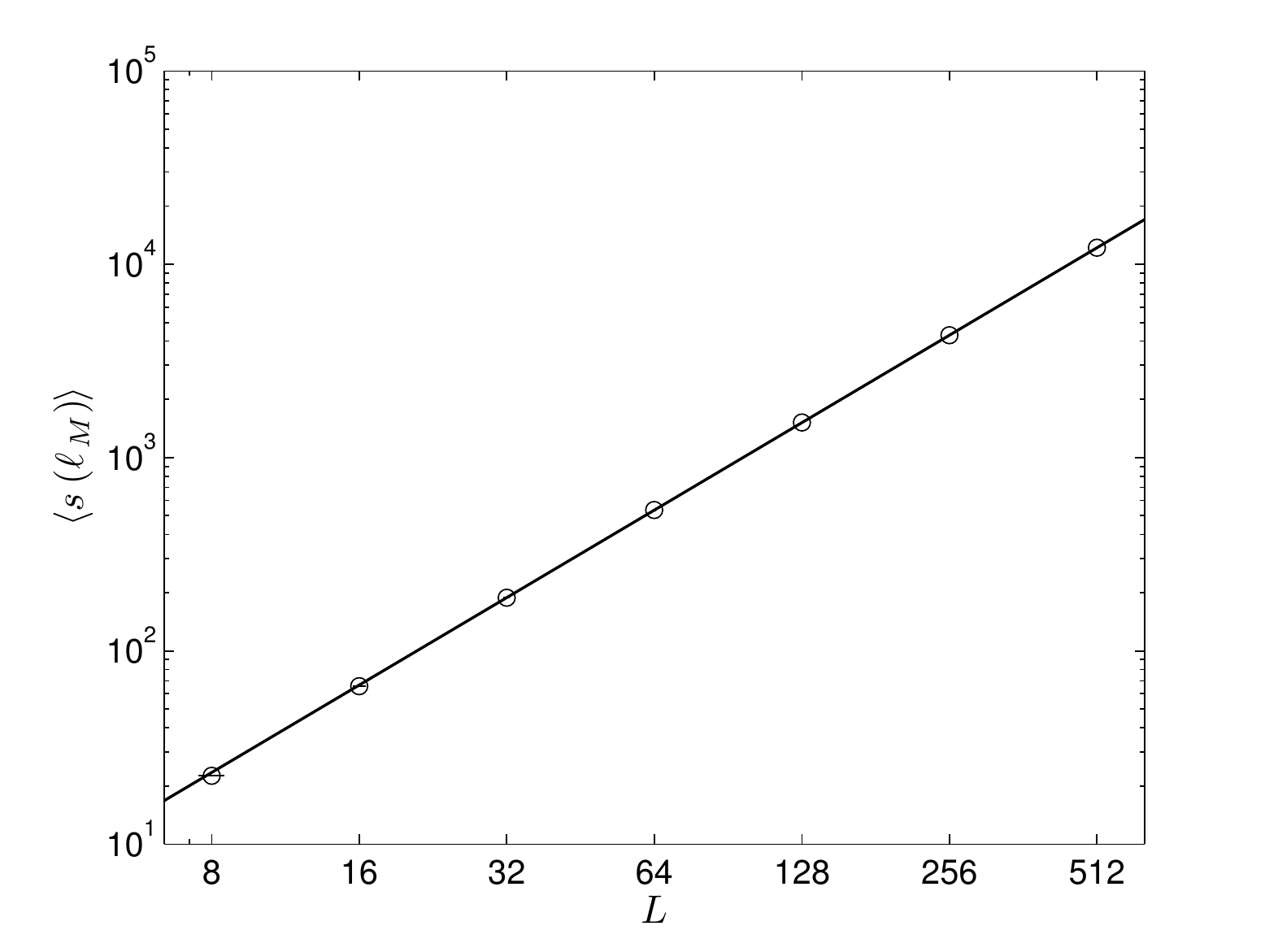} 
 \caption{ 
Finite size scaling of the length of longest loop, $s_M$ of the ground state of the RK point of square lattice QDM  ($\ket{\Psi_\TC^{e}}$). The best power-law fit gives $D_f = 1.502\pm0.002$, which is consistent with the fractal dimension of contour loops of a Gaussian height model.
}
\label{fig_sqRK_svL}
\end{figure}

For loops approaching a length $L^2$, the longest length scale of the lattice, we see a clear deviation from the power law behavior. Figure \ref{fig_sqRK_ll} shows the distribution of the longest loop; this displays the broad distribution over many length scales that is characteristic of a loop condensate. The location of the peak of $P(s_M)$ scales with system size, but not with an integral power (e.g., $L^0$, $L^1$, or $L^2$), which would be expected for dilute loop or symmetry broken phases, and therefore these spanning loops have a fractal dimensionality.
Figure \ref{fig_sqRK_svL} shows the finite size scaling of the fractal loop: the best fit power law gives $D_f = 1.502+\pm0.002$, which is clearly consistent with the universal height model fractal dimension prediction of $3/2$. 

These calculated geometric exponents provide quantitative confirmations of the height model description of the square lattice QDM. In contrast with the local correlation functions (for which the value of $K$ must be chosen to match the form of the height model correlations to the computed dimer model correlations functions), the universality of the fractal dimension of contour loops means that this quantitative agreement does not require tuning any parameters. Additionally we note that the fractal dimension of the transition loop distinguishes the critical liquid RK point from the gapped liquid of the triangular QDM, for which we have determined the fractal dimension to be $7/4$~\cite{Herdman2011}.

%%%%%%%%%%%%%%%%%%%%%%%%%
%Other loop condensates on the square lattice
%%%%%%%%%%%%%%%%%%%%%%%%%
\section{Other loop condensates on the square lattice}

%%%%%%%%%%%%
%Toric Code
\subsection{Toric code} 

\begin{figure}[] 
\centering
\includegraphics[width=3.5in]{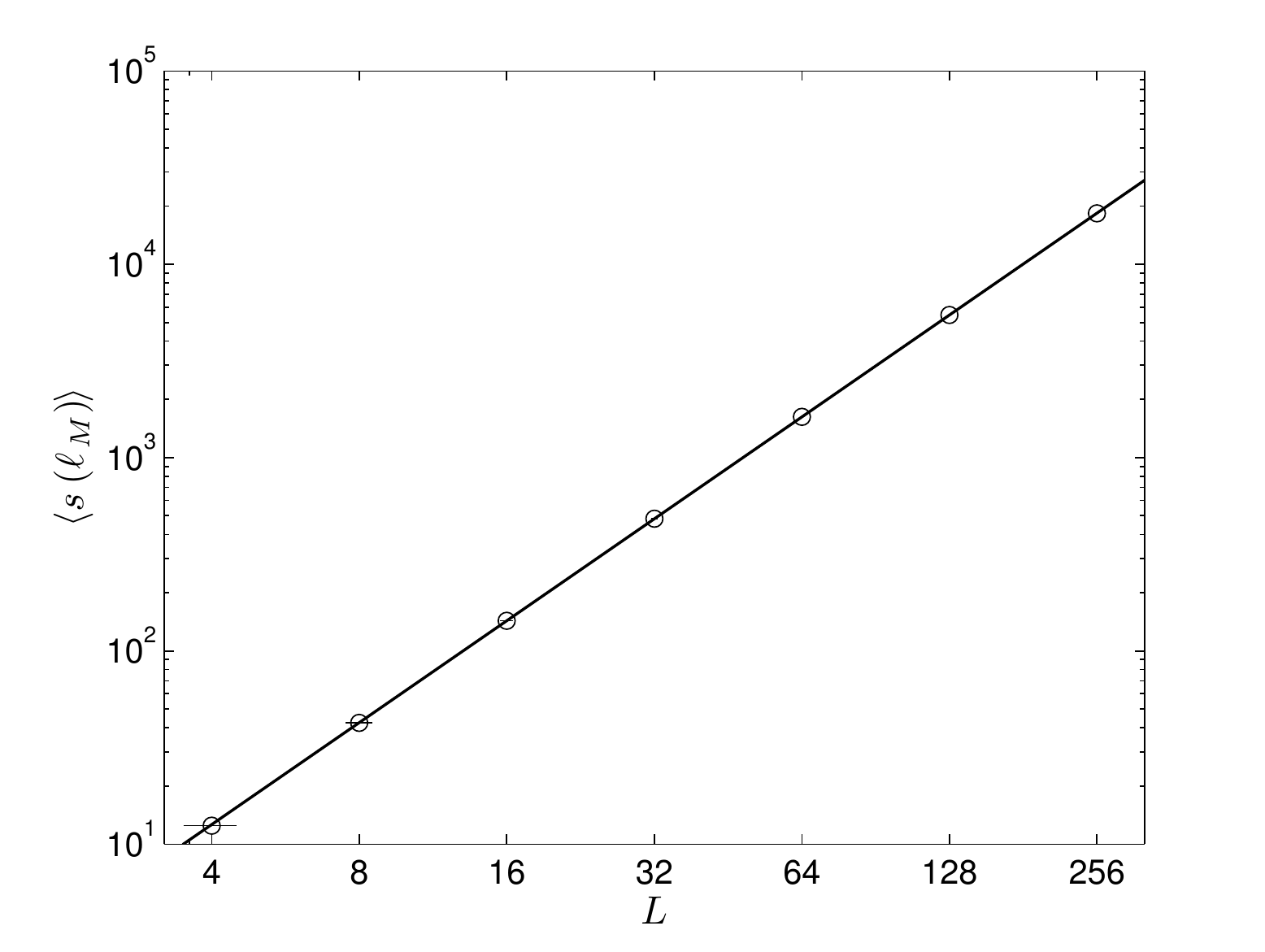} 
 \caption{ 
Finite size scaling of the longest loop in toric code ground state ($\ket{\Psi_\TC^{e}}$). We find $D_f = 1.7502\pm0.0002$, consistent with the universality of the fractal dimension of the gapped $Z_2$ liquid.
}
\label{fig_TC_OGT_fss}
\end{figure}

The ground state of the toric code, $\ket{\Psi_\TC^{e}}$, may be described as a loop condensate on the square lattice~\cite{Kitaev2003}. This gapped liquid phase has exponentially decaying spin-spin correlation functions. The ground-state subspace is that of closed loop coverings of the square lattice, with the exception that two loops may meet at a vertex. Since $\ket{\Psi_\TC^{e}}$ is an equal superposition of all loop coverings, we may relate it to the $O(1)$ loop model, by choosing a resolution of the four-loop vertices. Here we choose the orientation of the loops such that loops do not cross at an $n_v=4$ vertex (see Fig. \ref{fig_sq_constraint}). In Fig. \ref{fig_TC_OGT_fss}, the finite size scaling of the largest loop of the toric code wave function is plotted. From this, we have extracted the fractal dimension to be $D_f = 1.7502\pm0.0002$, which agrees with the known value for the $O(1)$ model, $D_f=7/4$~\cite{Duplantier1989}. We have also computed the fractal dimension for the odd Ising gauge theory described by $\ket{\Psi_\TC^{o}}$, and found that it is also $7/4$. Previous work has shown that the fractal dimension of $Z_2$ topological phases of the triangular lattice quantum dimer model and the honeycomb loop models both have fractal dimensions of $7/4$~\cite{Herdman2011,Troyer2008}. This suggests that the fractal dimension of $7/4$ is universal for the $Z_2$ topological liquid phases.

%%%%%%%%%%%%%
%Fully packed loop model
\subsection{Fully packed loop model} 

\begin{figure}[] 
\centering
 \includegraphics[width=3.5in]{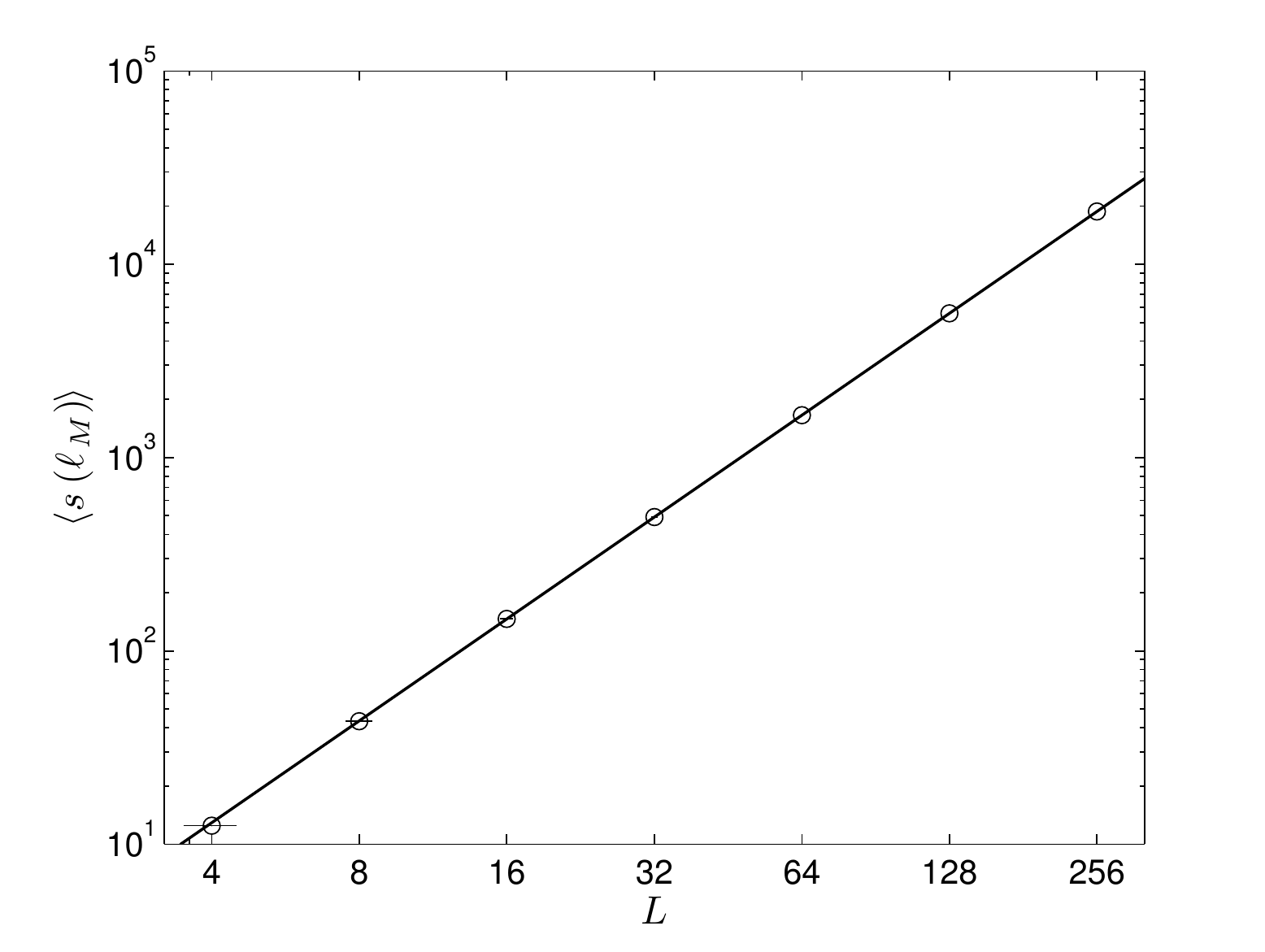} 
 \caption{ 
Finite size scaling of the length of the longest loop of the fully packed loop model that describes the 2-dimer constrained RK wave function, $\ket{\Psi_\RK^{2}}$. The best power-law fit gives $D_f = 1.7501\pm0.0002$, which is consistent with the fractal dimension of the effective height model.
}
\label{fig_FPLM_svL}
\end{figure}

The fully packed loop model is a critical liquid state with algebraically decaying loop-loop correlation functions~\cite{Kondev1997,Jacobsen1998}.
Just as with the RK point of the square lattice QDM there is also a phenomenological mapping of this model, which describes the RK wave function with a 2-dimer constraint at each vertex, $\ket{\Psi_\RK^{2}}$,  to a height model~\cite{Jacobsen1998,Kondev1997} . Similar to the dimer model case, the loops are oriented and act as the contour loops of the height field. However, in this case the effective action is a Gaussian action augmented by an additional term that is marginal and therefore changes the critical exponents~\cite{Kondev1997,Jacobsen1998,Nienhuis1984}. In this situation, the universality of the $D_f=3/2$ for a pure Gaussian model does not apply. Indeed, for the fully packed loop model, the height model mapping predicts a fractal dimension of $D_f=7/4$. We have confirmed this with directed loop Monte Carlo calculations of $\ket{\Psi_\RK^{2}}$ as summarized in \figref{fig_FPLM_svL}; a power-law fit yields $D_f = 1.7501\pm0.0002.$ While the  fractal dimension does not distinguish the fully packed loop condensate from the gapped $Z_2$ loop condensed phase, these two phases are easily distinguished by the loop-loop correlation function. In particular, the loop-loop correlation function displays a power law in the critical fully packed loop model, while it vanishes exponentially in the gapped liquid phase.

%%%%%%%%%%%%%%%%%%%%%%%%%
%Conclusions
%%%%%%%%%%%%%%%%%%%%%%%%%
\section{Conclusions} 

We have studied the geometric properties of the transition loops of ground state of the RK point of the square lattice QDM. The power-law distribution of loop lengths as well as the fractal dimensionality of loops that span the finite system indicate that this critical dimer liquid state is fundamentally described by a loop condensate. Our numerically computed geometric exponents agree with the predictions of the standard height model description, which gives a quantitative confirmation of the relevance of this phenomenological action. We have compared the loop condensate of the square lattice QDM with those of the toric code and fully packed loop models on the square lattice. The fractal dimension of $3/2$ of the square lattice QDM distinguishes it from the gapped $Z_2$ topologically ordered phase of the toric code on the square lattice and that of the QDM on the triangular lattice, both of which have fractal dimension of $7/4$. However the critical loop condensate of the fully packed loop model has the same fractal dimension as these gapped phases, so the current work demonstrates that these geometric exponents may not always distinguish loop condensed phases. Similar analyses of other quantum liquids phases may prove fruitful for characterizing other loop condensed phases. 

%%%%%%%%%%%%%%%%%%%%%%%%%
%Acknowledgements
%%%%%%%%%%%%%%%%%%%%%%%%%
\acknowledgements
This work was supported by NSF grant number PH4-0803429.

%----------------------------------------------------------------
%------ Bibliography
%----------------------------------------------------------------
\bibliographystyle{apsrev4-1}
\bibliography{/Users/Chris/Documents/Bibliographies/LoopModels-LCsqL}

\end{document}